\documentclass[a4,center,fleqn]{NAR}

\newenvironment{myenumerate2}{
\begin{enumerate}
 \setlength{\itemsep}{1pt}
\setlength{\parskip}{0pt}
 \setlength{\parsep}{0pt}}{\end{enumerate}
}

\newenvironment{myitemize}{
\begin{itemize}
 \setlength{\itemsep}{1pt}
 \setlength{\parskip}{0pt}
 \setlength{\parsep}{0pt}}{\end{itemize}
}

\copyrightyear{20XX}
\pubdate{31 July 20XX}
\pubyear{20XX}
\jvolume{XX}
\jissue{XX}

\begin{document}

%\title{Training-free DNA Sequence Complexity Indices Reveal Structure Beyond GC Content in Nucleosome Occupancy}

\title{Training-free Measures Based on Algorithmic Probability Identify High Nucleosome Occupancy in DNA Sequences}

\author{%
Hector Zenil\,$^{1,2,3}$\footnote{Corresponding: hector.zenil@algorithmicnaturelab.org}
and Peter Minary\,$^2$%
}

\address{%
$^{1}$Algorithmic Dynamics Lab, Unit of Computational Medicine, SciLifeLab, Centre for Molecular Medicine, Karolinska Institute, Stockholm, Sweden
and
$^{2}$Department of Computer Science, University of Oxford, Oxford, UK
$^{3}$Algorithmic Nature Group, LABORES for the Natural and Digital Sciences, Paris, France.}

\history{%
Received XXX X, 201X;
Revised XXX X, 201X;
Accepted XXX X, 201X}

\maketitle

\begin{abstract}
We introduce and study a set of training-free methods of information-theoretic and algorithmic complexity nature applied to DNA sequences
to identify their potential capabilities to determine nucleosomal binding sites. 
We test our measures on well-studied genomic sequences of different sizes drawn from different sources. The measures reveal the known in vivo versus in vitro predictive discrepancies and uncover their potential to pinpoint (high) nucleosome occupancy. We explore different possible signals within and beyond the nucleosome length and find that complexity indices are informative of nucleosome occupancy. We compare against the gold standard (Kaplan model) and find similar and complementary results with the main difference that our sequence complexity approach. For example, for high occupancy, complexity-based scores outperform the Kaplan model for predicting binding representing a significant advancement in predicting highest nucleosome occupancy following a training-free approach.
\end{abstract}

\section{The challenge of predicting nucleosome organisation}

DNA in the cell is organised into a compact form, called chromatin~\cite{tanmoy}. One level of chromatin organisation consists in DNA wrapped around histone proteins, forming nucleosomes~\cite{reece}. A nucleosome is a basic unit of DNA packaging. Depending on the context, nucleosomes can inhibit or facilitate transcription factor binding and are thus a very active area of research. The location of low nucleosomal occupancy is key to understanding active regulatory elements and genetic regulation that is not directly encoded in the genome but rather in a structural layer of information.

The structural organisation of DNA in the chromosomes is widely known to be heavily driven by GC content~\cite{Yeast}, notwithstanding that $k$-mer approaches have been discovered to increase predictive power~\cite{struhl,changu,schep}. Indeed, local and short-range signals carried by DNA sequence `motifs' or fingertips have been found to be able to determine a good fraction of the structural (and thus functional) properties of DNA, such as nucleosome occupancy, with significant differences for in vivo vs. in vitro data~\cite{kaplan}. 

Despite intensive analysis of the statistical correspondence between in vitro and in vivo positioning, there is a lack of consensus as to the degree to which the nucleosome landscape is intrinsically specified by the DNA sequence~\cite{gracey}, as well as in regards to the apparently profound difference in dependence in vitro vs. in vivo.

Here, we consider a set of algorithmic and information-theoretic complexity measures to help unveil how much of the information encoded in a sequence in the context of the nucleosome landscape can be recovered from training-free information-content and algorithmic complexity measures, i.e. with no previous knowledge such as informative $k$-mers. Nucleosome location is an ideal test case to probe how informative sequence-based indices of complexity can be in determining structural (and thus some functional) properties of genomic DNA, and how much these measures can both reveal and encode.

\subsection{Information-theoretic approaches to Genomic Profiling}

Previous applications of measures based upon algorithmic complexity include experiments on the evaluation of lossless compression lengths of sets of genomes~\cite{rivals,vitanyi2}, and, more recently, in~\cite{pratas},
demonstrating applications of algorithmic complexity to DNA sequences. In a landmark paper in the area, a measure of algorithmic mutual information was introduced to distinguish sequence similarities by way of minimal encodings and lossless compression algorithms in which a mitochondrial phylogenetic tree that conformed to the evolutionary history of known species was reconstructed~\cite{vitanyi,vitanyi2}. 

\newpage 

However, most of these approaches have either been purely theoretical or have been effectively reduced to applications or variations of Shannon entropy~\cite{utro} rather than of algorithmic complexity, because popular implementations of lossless compression algorithms are actually closer to Shannon entropy than to algorithmic complexity~\cite{emergence,smalldata}. 

In certain cases, some control tests have been missing. For example, in the comparison of the similarity distances of different animal genomes~\cite{vitanyi,vitanyi2} based on lossless compression, GC content (counting every G and C in the sequence) can reconstruct an   animal phylogenetic tree as accurate as the one produced~\citep{pozzoli}. This is because two species that are close to each other evolutionarily will also have similar GC content. 

Species close to each other will have similar DNA sequence entropy values, allowing lossless compression algorithms to compress statistical regularities of genomes of related species with similar compression rates. Here we intend to go beyond previous attempts, in breadth as well as depth, using better-grounded algorithmic measures and more biologically relevant test cases. Indeed, the GC content of every species can be mapped to a single point--and its complement-- on a Bernoulli-shaped curve of Shannon entropy corresponding to the count of G or C vs. A or T. 

\subsection{Current sequence-based prediction methods}

While the calculation of GC content is extremely simple, the reasons behind its ability to predict the structural properties of DNA are not completely understood~\cite{Yeast,galtier}. For example, it has been shown that low GC content can explain low occupancy, but high GC content can mean either high or low occupancy~\cite{minary}. But how much GC content alone encodes nucleosome position, given that DNA (and thus GC content) encodes much more than chromatin structure, is a topic of interest. The same DNA sequences are constrained within functional/evolutionary trajectories such as protein coding vs. non-coding and regulatory vs. non regulatory, among others. The in vitro and in vivo discrepancy can be explained in the same terms, with other factors such as chromatin remodellers and transcription factors affecting nucleosome organisation differently in vitro vs. in vivo.

Current algorithms that build upon while probing beyond GC content have been largely influenced by sequence motif  ~\cite{cui,schep} and dinucleotide models ~\cite{trifonov}---and to a lesser degree by $k$-mers~\cite{changu}, and thus are not training- free, and are the result of years of experimental research.

\subsection{The Dinucleotide Wedge Model}

The formulation of models of DNA bending was initially prompted by a recognition that DNA must be bent for packaging into nucleosomes, and that bending would be an informative index of nucleosome occupancy. Various dinucleotide models can account reasonably well for the intrinsic bending observed in different sets of sequences, especially those containing A-tracts~\cite{kanhere}.

The \textit{Wedge model}~\cite{ulanovsky} suggests that bending is the result of driving a wedge between adjacent base pairs at various positions in the DNA. The model assumes that bending can be explained by wedge properties attributed solely to an AA dinucleotide (8.7 degrees for each AA). No current model provides a completely accurate explanation of the physical properties of DNA such as bending~\cite{burkhoff}, but the Wedge model (like the more basic Junction model, which is less suitable for short sequences and less general~\cite{donald}) reasonably predicts the bending of many DNA sequences~\cite{sinden}. Although it has been suggested that trinucleotide models may make for greater accuracy in explaining DNA curvature in some sequences, dinucleotide models remain the most effective~\cite{kanhere}.

\subsection{The Kaplan Model}

Kaplan et al. established a probabilistic model to characterise the possibility that one DNA sequence may be occupied by a nucleosome~\cite{kaplan}. They constructed a nucleosome-DNA interaction model and used a hidden Markov model (HMM) to obtain a probability score. The model is based mainly on a 10-bp sequence periodicity that indicates the probability of any base pair being covered by a nucleosome. 
The Kaplan model is considered the most accurate, and is the gold standard for predicting in vitro nucleosome occupancy. However, previous approaches, including Segal's~\cite{segal} and Kaplan's~\cite{kaplan}, require extensive (pre-)training. In contrast, all measures considered in our approach are training-free. The model of Kaplan et al. is considered the gold standard for comparison purposes.

%All $k$-nucleotide models, including that of Segal and Kaplan et al., are based upon knowledge-based sequence motifs and are thereby dependent on certain previously learned patterns. They can only account for local curvature and local predictions, not longer range correlations. Perhaps the fact that $k$-nucleotide models for $k>2$ have not been proven to provide a significant advantage over $k=2$ has led researchers to disregard longer range signals across DNA sequences involved in both DNA curvature and nucleosome occupancy~\cite{heijden}. To date, these models, including that of Kaplan~\cite{kaplan} (which considers up to $k=5$) 

\begin{figure*}
  \centering
\includegraphics[width=11cm]{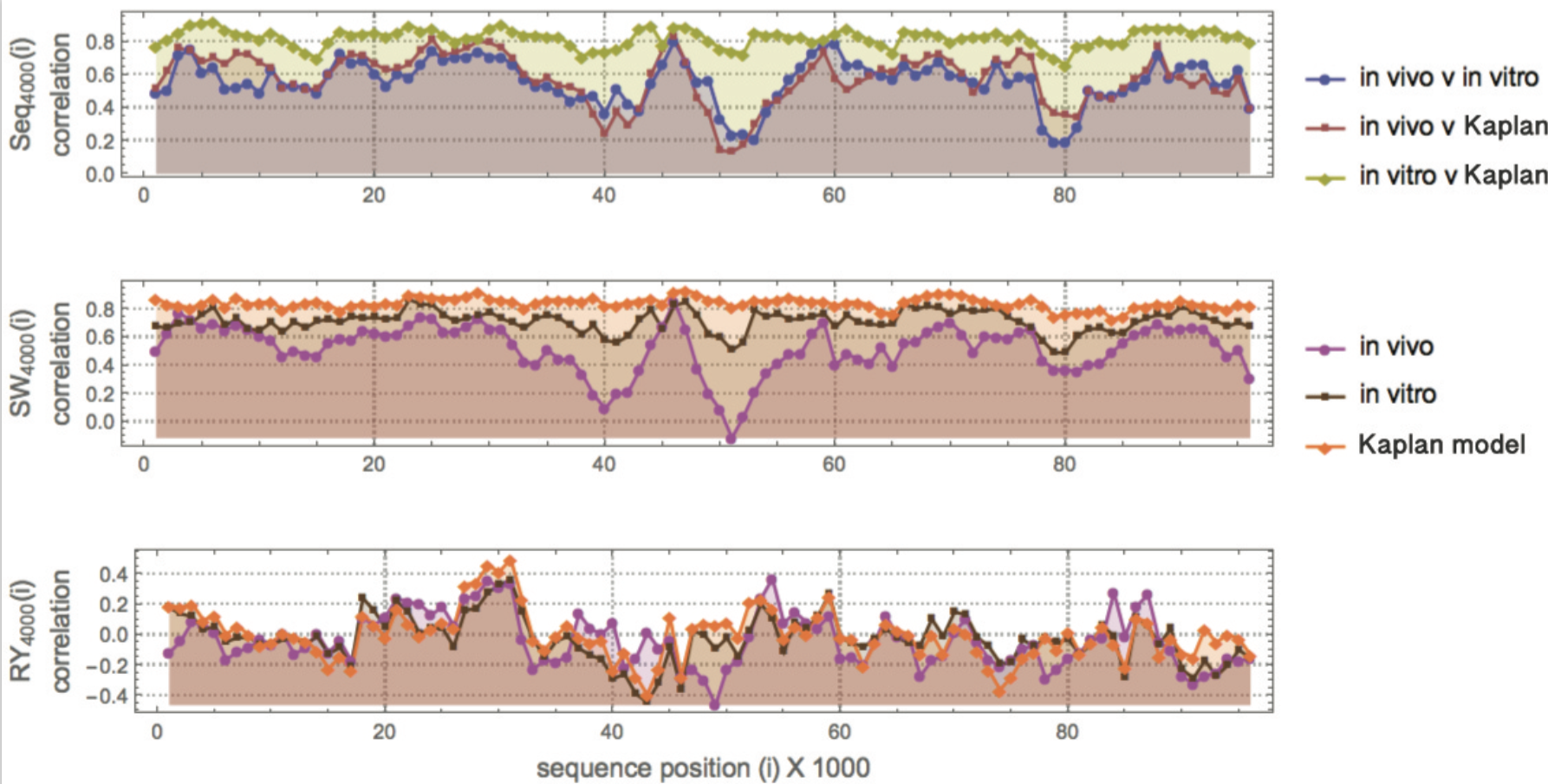}
 \caption{Top: Correlation values of nucleosome occupancy in the 14th Yeast chromosome (experimentally validated vs. Kaplan model) on a sliding window of length 4K nt for both in vitro and in vivo data against different measures/signals: the occupancy predicting Kaplan model (clearly better for in vitro). Middle: SW is simply GC written as SW in contrast to RY (which is not AT). Calculated correlation values are highly correlated to the Kaplan model but are poor at explaining in vivo occupancy data. Bottom: The RY DNA transformation, an orthogonal signal to SW (and thus to GC content) whose values report a non-negligible max-min correlation, suggesting that the mixing of AT and GC carries some information about nucleosome occupancy (even if weaker than GC content), with in vivo values showing the greatest correlation values, unlike SW/GC, and thus possibly neglected in predictive models (such as Kaplan's).}
  \label{plots}
\end{figure*}

\section{Methods}
\label{methodology}

To study the extent of some signals in the determination of nucleosome occupancy, we applied some basic transformations to the original genomic DNA sequence. The SW transformation substitutes G and C for S (Strong interaction), and A and T for W (Weak interaction). The RY transformation substitutes A and G for R (puRines), and C and T for Y (pYrimidines).

\subsection{Complexity-based genomic profiling}

In what follows, we generate a function score $f_c$ for every complexity measure $c$ (detailed descriptions in the Sup. Mat.) by applying each measure to a sliding window of length 147 nucleotides (nts) across a 20K and 100K base pair (bps) DNA sequence from Yeast chromosome 14~\cite{Yeast}. At every position of the sliding window, we get a function score for every complexity index $c$ applied to the sequence of interest used to compare in vivo and in vitro occupancies.

The following measures (followed in parentheses by the names we use to refer to them throughout the text) are introduced here. Among the measures considered are entropy-based ones (see Supplementary Material for exact definitions): 

\begin{myitemize}
\item Shannon entropy with uniform probability distribution.
\item Entropy rate with uniform probability distribution.
\item Lossless compression (Compress)
\end{myitemize}

A set of measures of algorithmic complexity (see Supplementary Material for exact definitions):

\begin{itemize}
\item Coding Theorem Method (CTM) as an estimator of algorithmic randomness by way of algorithmic probability via the algorithmic Coding theorem (see Supplementary Material) relating causal content and classical probability~\cite{delahayezenil,zenild5}.
\item Logical Depth (LD) as a BDM-based (see below) estimation of logical depth~\cite{bennett}, a measure of sophistication that assigns both algorithmically simple and algorithmically random sequences shallow depth, and everything else higher complexity, believed to be related to biological evolution~\cite{bennett2,alife}.
\end{itemize}

And a hybrid measure of complexity combining local approximations of algorithmic complexity by CTM and global estimations of (block) Shannon entropy (see Sup. Mat. for exact definitions):

\begin{itemize}
\item The Block Decomposition Method (BDM) that approximates Shannon entropy---up to a logarithmic term---for long sequences, but Kolmogorov-Chaitin complexity otherwise, as in the case of short nucleotides~\cite{bdm}.
\end{itemize}

We list lossless compression under information-theoretic measures and not under algorithmic complexity measures, because popular implementations of lossless compression algorithms such as Compress and all those based on Lempel?Ziv?Welch (LZ or LZW) as well as derived algorithms (ZIP, GZIP, PNG, etc.) are actually entropy estimators~\cite{emergence,smalldata,bdm}.  

BDM allows us to expand the range of application of both CTM and LD to longer sequences by using Shannon entropy. However, if sequences are divided into short enough subsequences (of 12 nucleotides), we can apply CTM and avoid any trivial connection to Shannon entropy and thus to GC content.

Briefly, to estimate the \textit{algorithmic probability}~\cite{solomonoff,levin}---on which the measure BDM is based---of a DNA sequence (e.g. the sliding window of length 147 nucleoides or nt), we produce an empirical distribution~\cite{delahayezenil,zenild5} to compare with by running a sample of 325\,433\,427\,739 Turing machines with 2 states and 4 symbols (which is also the number of nucleotide types in a DNA sequence) with empty input. If a DNA sequence is algorithmically random, then very few computer programs (Turing machines) will produce it, but if it has a regularity, either statistical or algorithmic, then there is a high probability of its being produced. Producing approximations to algorithmic probability provides approximations to algorithmic complexity by way of the so-called \textit{algorithmic Coding Theorem}~\cite{levin,delahayezenil,zenild5}. Because the procedure is computationally expensive (and ultimately uncomputable), only the full set of strings of up to 12 bits was produced, and thus direct values can be given only to DNA sequences of up to 12 digits (binary for RY and SW and quaternary for full-alphabet DNA sequences).

\section{Results}

Table~\ref{infomeasures} (Sup. Mat.) shows the in vitro nucleosome occupancy dependence on GC content, with a correlation of 0.684 (similar to that reported by Kaplan~\cite{kaplan}) for the well-studied 20K bp genomic region (187K -- 207K) of Yeast Chromosome 14, exactly as was done and reported in~\cite{segal} using their data with no sliding window but on full sequences.
%The range of in vitro occupancies generally increases as a function of increasing GC content. However, at low GC content there is weak nucleosome binding, whereas at high GC content nucleosome occupancy can be moderate or high. 
Knowledge-based methods dependent on observed sequence motifs~\cite{lee} are computationally cost-effective alternatives for predicting genome-wide nucleosome occupancy. However, they are trained on experimental statistical data and are not able to predict anything that has not been observed before. They also require context, as it may not be sufficient to consider only short sequence motifs such as dinucleotides~\cite{kanhere,kaplan}. 

More recently, deep machine learning techniques have been applied to DNA accessibility related to chromatin and nucleosome occupancy~\cite{kelley}. However, these techniques require a huge volume of data for training if they are to predict just a small fraction of data with marginally improved accuracy, as compared to more traditional approaches based on $k$-mers, and they have not shed new light on the sequence dependence of occupancy. 

Here we test the ability of a general set of measures, statistical and algorithmic, to be informative about nucleosome occupancy and/or about the relationship between the affinity of nucleosomes with certain sequences and their complexities.

\subsection{Complexity-based indices}

Fig.~\ref{plots} shows the correlations between in vivo, in vitro data, and the Kaplan model. In contrast, the SW transformation captures GC content, which clearly drives most of the nucleosome occupancy, but the correlation with the RY transformation that loses all GC content is very interesting. While significantly lower, it does exist and indicates a signal not contained in the GC content alone, as verified in Fig.~\ref{topallplots}. 

In Table~\ref{infomeasures} (Sup. Mat.), we report the correlation values found between experimental nucleosome occupancy data and ab initio training-free complexity measures. BDM alone explains more than any other index, including GC content in vivo, and unlike all other measures LD is negatively correlated, as theoretically expected~\cite{computability} and numerically achieved~\cite{bdm}, it being a measure that assigns low logical depth to high algorithmic randomness, with high algorithmic randomness implying high entropy (but not the converse).

Entropy alone does not capture all the GC signals, which means that there is more structure in the distributions of Gs and Cs beyond the GC content alone. However, entropy does capture GC content in vivo, suggesting that local nucleotide arrangements (for example, sequence motifs) have a greater impact on in vivo prediction. Compared to entropy, BDM displays a higher correlation with in vivo nucleosome occupancy, thereby suggesting more internal structure than is captured by GC content and entropy alone, that is, sequence structure that displays no statistical regularities but is possibly algorithmic in nature.

\subsection{Model curvature vs. complexity indices}

\text{ } The dinucleotide model incorporates knowledge regarding sequence motifs that are known to have specific natural curvature properties, and adds to the knowledge and predictive power that GC content alone offers. 

Using the Wedge dinucleotide model we first estimated the predicted curvature on a set of 20 artificially generated sequences (Table~\ref{sequences} (Sup. Mat.)) with different statistical properties, in order to identify possibly informative information-theoretic and algorithmic indices. As shown in Table~\ref{heatmap} (Sup. Mat.), we found all measures negatively correlated to the curvature modelled, except for LD, which displays a positive correlation--and the highest in absolute value--compared to all the others. Since BDM negatively correlates with curvature, it is expected that the minima may identify nucleosome positions (see next subsection). 

An interesting observation--see Table~\ref{heatmap} (Sup. Mat.)--concerning the correlation values between artificially generated DNA sequences and DNA structural curvature according to the Wedge nucleotide model: all values are negatively correlated, but curvature as predicted by the model positively correlates with LD, in exact inverse fashion vis-\`a-vis the correlation values reported in Table~\ref{infomeasures} (Sup. Mat.). This is consonant with the theoretically predicted relation between algorithmic complexity and logical depth~\cite{bdm}. All other measures (except for LD) behave similarly to BDM.

The results in Tables~\ref{infomeasures} and~\ref{heatmap} (Sup. Mat.) imply that for all measures, extrema values may be indicative of high nucleosome occupancy. In the next section we explore whether extrema of these measures are also informative about nucleosome location.

\subsection{Nucleosome Dyad and Centre Location Test}
\label{dyadcentre}

The positioning and occupancy of nucleosomes are closely related. Nucleosome positioning is the distribution of individual nucleosomes along the DNA sequence and can be described by the location of a single reference point on the nucleosome, such as its dyad of symmetry~\cite{klug}. Nucleosome occupancy, on the other hand, is a measure of the probability that a certain DNA region is wrapped around a histone octamer.

Here we have taken a set of sequences that are, to our knowledge, among the most studied in the context of nucleosome research. Their structural properties have been experimentally validated, making them ideal for testing any measure on, and they have also been used in other studies.

\begin{figure*}
  \centering
  \includegraphics[width=11cm]{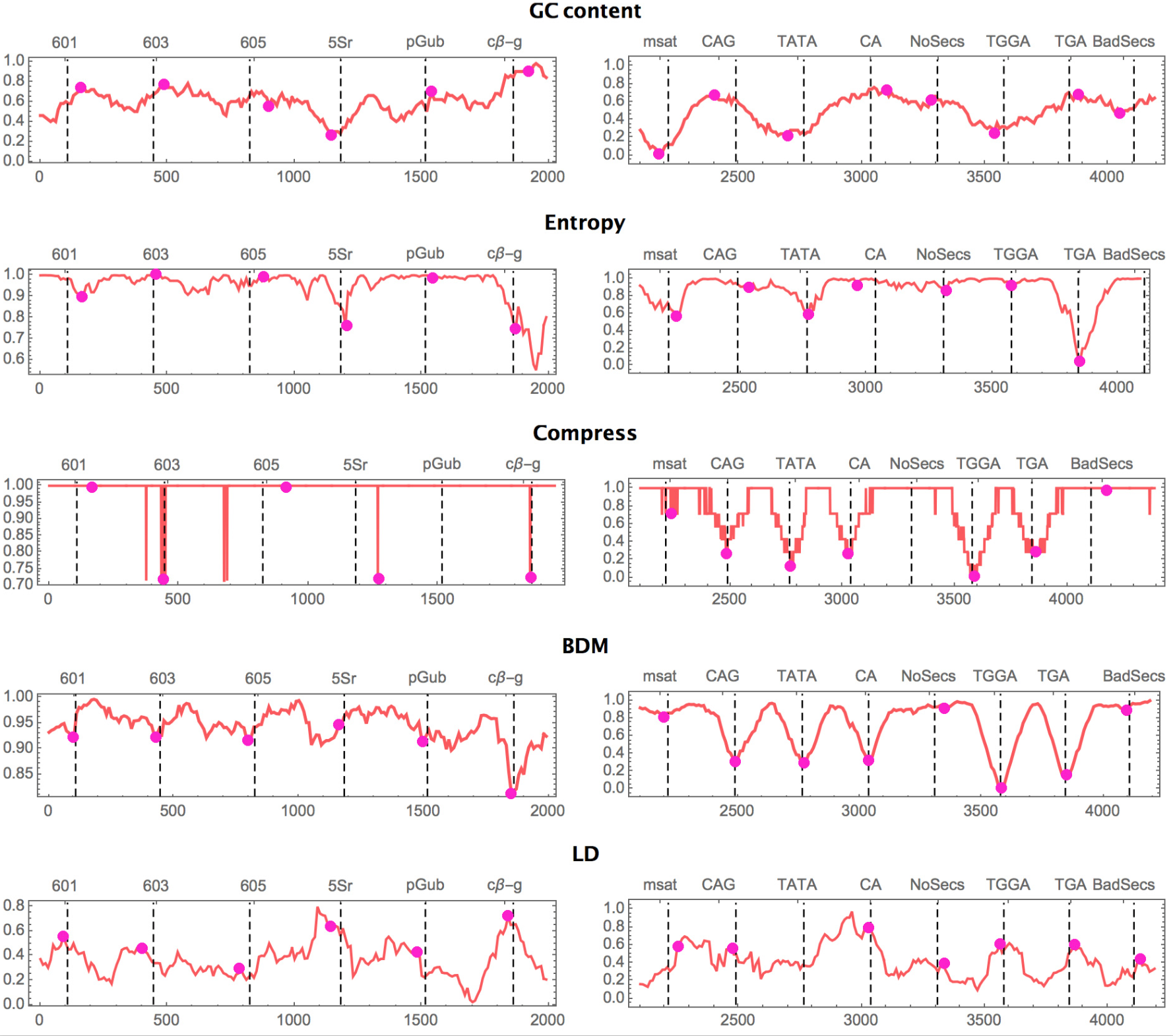}
\caption{Nucleosome centre location according to five indices on 14 highly-studied and experimentally validated nucleosomes in Yeast (source and full sequences listed in Table~\ref{nucleosome} in the Supplementary Material) intercalated with another 16 pseudo-random DNA segments of 147 nts with the same average GC content as one of the immediate neighbouring legitimate nucleosomal sequences to erase any GC content difference on purpose. Values are normalised between 0 and 1 and they were smoothed by taking one data point per 10. The $y$ axis scales differ between the left and right panels for ease of illustration only. Experimentally known nucleosome centres (called dyads) are marked with dashed lines and located centres according to each measure are marked with a magenta circle. Panels on the right for which no dyad is known have their centre estimated by the centre of the nucleosomal sequence. By design GC content performs poorly, but entropy recovers the signal ab initio. Centre predictions were called based on the local (147 nt window) minimum; only GC Content was called based on  either the local minimum or local maximum (min/max), thus giving it an extra edge. LD centre calls were made to the local maximum. Values for the best performer index, BDM, are reported in Table~\ref{nucleosomedistances}. Due to boundary effects, the authors highly recommend not to apply the Kaplan model on less than 5000 bp which would have forced us to flank our sequences with 10 times more pseudo-random sequences. The Kaplan model thus is unsuitable to deal with to short sequences and was therefore not included given that according to the authors any result would have been not statistically significant. We know, however, that the Kaplan model heavily relies on GC-content and $K$-mers so it is expected to be fooled after flanking with sequences of similar GC content.}\label{3a}
\end{figure*}

\begin{figure}[htpb]
  \centering
    \includegraphics[width=7cm]{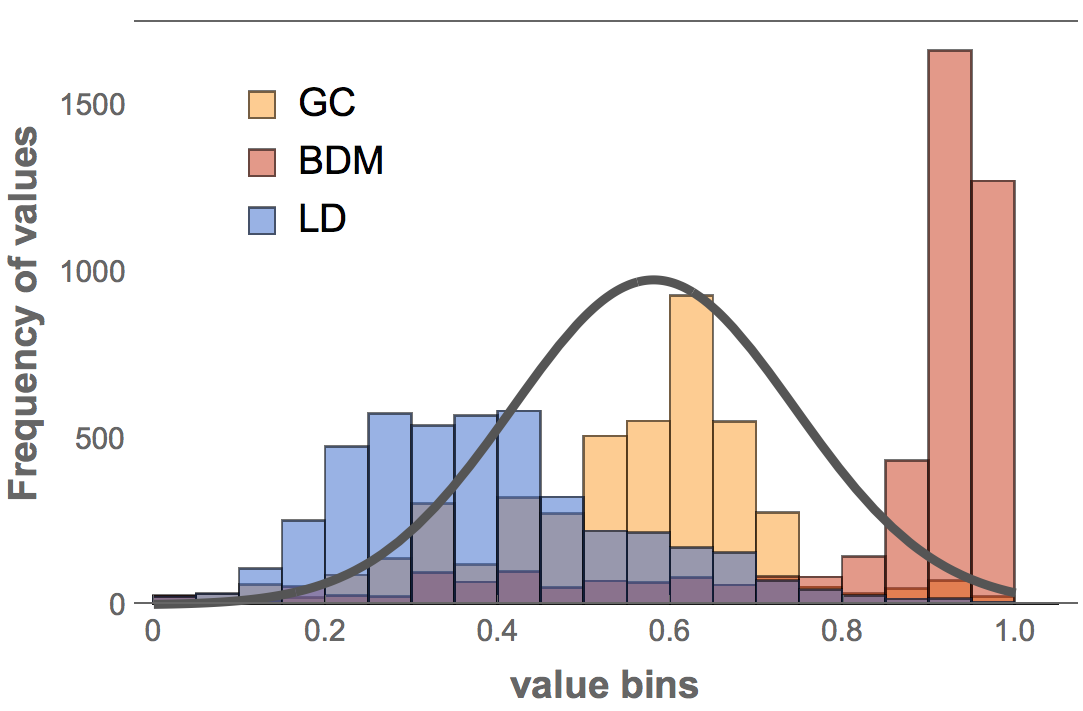}
\caption{Histogram of values taken from the experiments reported in Fig.~\ref{3a} (thus normalised between 0 and 1) demonstrating how BDM and LD are the most removed from a normal distribution, unlike GC content that distributes values closer to normal, as expected, given the nature of the experiment the purpose of which was to dissimulate GC content and see if other measures can overcome it, thus identifying alternative measures. A normal probability distribution is plotted in black colour with mean and std deviation estimated from the GC content values for comparison purposes. BDM carries the strongest signal, followed by LD skewed in the opposite direction, with both peaking closer to the nucleosome centres than GC content. On the $x$-axis are complexity values arranged in bins of 1000, as reported in Fig.~\ref{3a}.}
  \label{3b}
\end{figure}

Fig.~\ref{3a} shows the location capabilities of algorithmic indices for nucleosome dyad and centre location when nucleosomal regions are placed against a background of (pseudo-) randomly generated DNA sequences with the same average GC content as the immediate left legitimate neighbour. As illustrated, BDM outperforms all methods in accuracy (Fig.~\ref{3a} and Table~\ref{nucleosomedistances} (Sup. Mat.)) and in signal strength (Fig.~\ref{3b}). The results indicate an emerging trend. For algorithmic and information-theoretic measures the minimum is correlated with nucleosomal centre location, except for LD (which is weakly negatively correlated). The problematic case happens to be GC content, which is sometimes max and sometimes min the one closest to the nucleosomal centre, as expected, given that the experiment is designed to 
conceal any GC content differences by inserting 
spurious sequences around it with the same GC content as the nucleosomal regions, and so is expected to be weakly indicative in either direction, if at all, as is the case here.

The results for BDM and LD suggest that the first 4 nucleosomal DNA sequences, of which 3 are clones, display greater algorithmic randomness (BDM) than the statistically pseudo-randomly generated background (surrounding each legitimate sequence) designed to 
erase any GC content difference, while all other nucleosomes are of significantly lower algorithmic randomness (BDM) and mixed (both high and low) structural complexity (LD). Structural complexity in the context of LD means sequences that are deep in computational content, that is, they are neither random nor trivial and they require computational work (the segments can only be generated by a slow computer program). The same robust results were obtained after several replications with different pseudo-random backgrounds. Moreover, the signal produced by similar nucleosomes with strong properties~\cite{gaykalova}, such as clones 601, 603 and 605, had similar shapes and convexity. The results suggest that algorithmic and information-theoretic measures can recover a strong signal and can complement GC content and $K$-mer training at finding nucleosome positions and nucleosomal centres.

Fig.~\ref{3b} shows the strength of the BDM signal at indicating the nucleosome centres based on the local minima for all measures (except LD) or the min/max value for GC content. The signal-to-noise ratio is much stronger for BDM. LD is shifted in the opposite direction (to BDM), consistent with the theoretical expectation (what is highly random for BDM is shallow for LD) (see Sup. Mat.).

Both BDM and LD spike at nucleosome positions stronger than GC content on a random DNA background with the same GC content, and perform better than entropy and compression. BDM is informative about every dyad or centre of nucleosomes, with 8 out of the 14 predicted within 1 to 7 nts distance. Unlike all other measures, LD performed better for the first half (left panel) of nucleosome centre locations than for the second half (right panel), suggesting that the nucleosomes of the first half may have greater structural organisation.

Table~\ref{nucleosomedistances} (Sup. Mat.) compares distances to the nucleosome centres as predicted without any training, with BDM outperforming GC content as shown in Fig.~\ref{3a}. The average distance between the predicted and the actual nucleosome centre is calculated to the closest local extreme (minima or maxima) for GC content and only minima for BDM (hence giving GC content an advantage) within a window of 73 bps from the actual centre (the experimentally known dyad, or the centre nucleotide when the dyad is not known).

In accordance with the results provided in Fig.~\ref{3a} and Table~\ref{infomeasures} (Sup. Mat.) the minima of BDM is informative for nucleosome position for the 14 test sequences whose natural curvature is a fit to the superhelix. The minima of BDM (maxima of LD) may thus also indicate  nucleosome location. This latter finding is supported by results in Table~\ref{heatmap} (Sup. Mat.).

Our results suggest that if some measures of complexity indicate occupancy nucleosomal regions where GC content is (purposely) falsified, the measures capture structural signals different from GC content such as $k$-mers accounting for less than 20\% of the accuracy of the Kaplan model (with the rest owing to GC content alone). However, the strong signal captured by some complexity measures and the marks found in signals complementary to GC content (RY content) suggest that these complexity measures are not only able to capture the usual markers, such as GC content with e.g. Shannon entropy alone, but also $k$-mer knowledge without any previous knowledge or training. Furthermore, the measures may be revealing signals complementary to GC content running along the DNA not revealed hitherto and requiring further examination.

\subsection{Informative Measures of High and Low Occupancy}
\label{informative}

To find the most informative measures of complexity $c$ we maximised the separation by taking only the sequences with the highest 2\% and lowest 0.2\% nucleosome occupancy from a 100K DNA segment for highest and lowest nucleosome occupancy values. There were 7701 high and 5649 low occupancy in vitro sequences, and 4332 high and 3989 low in vivo sequences. The starting and ending points of the 100K segment are 187K $-$ 40K and 207K $+$ 40K nts in the same 14th Yeast chromosome~\cite{Yeast,segal}, so 40K nts surrounding the original shorter 20K sequence first studied in this paper.

\begin{figure*}
  \centering
\includegraphics[width=13cm]{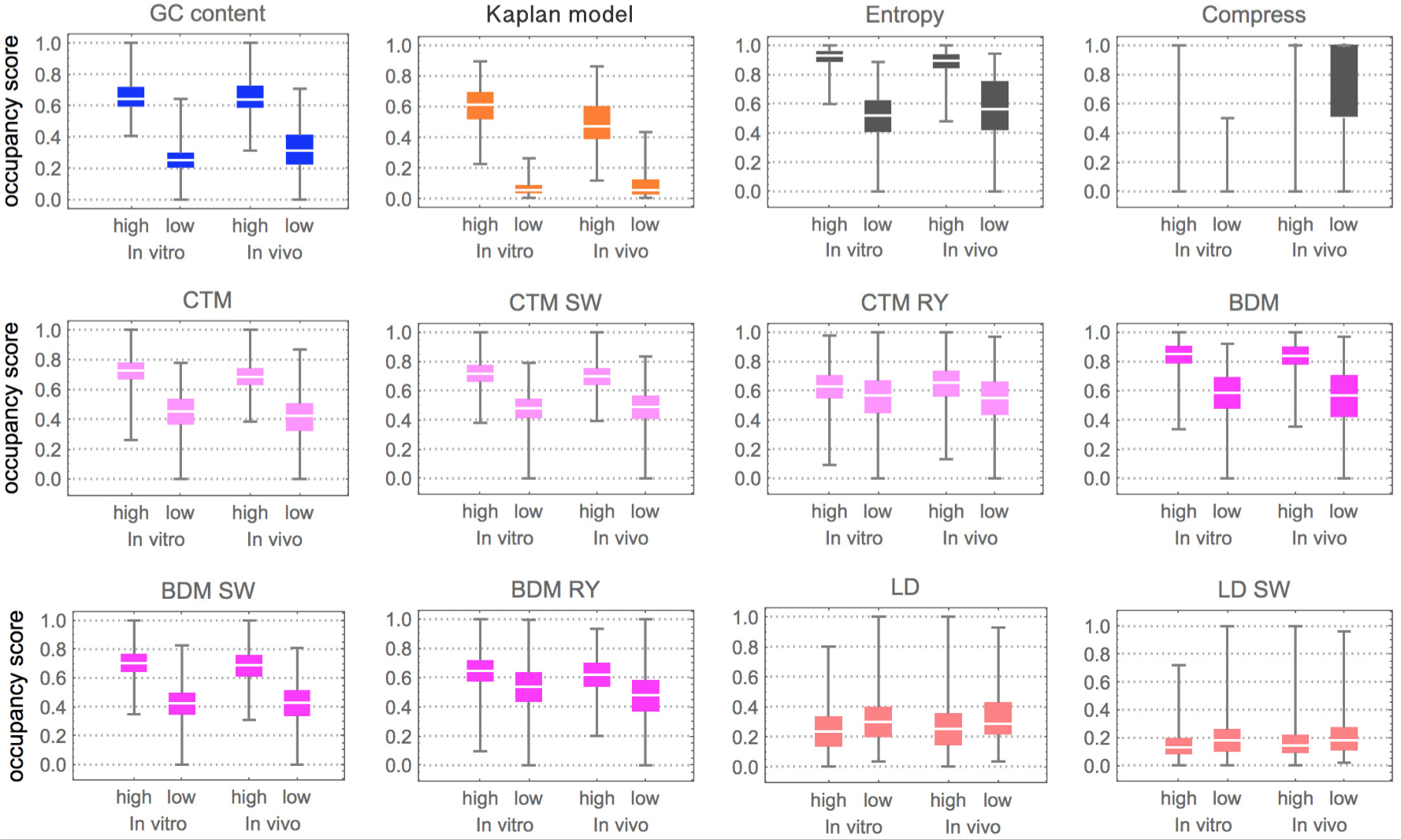}
\caption{Box plots of informative indices for top highest and bottom lowest occupancies on three regions of Yeast chromosome 14 of 100K bps, representing about 1\% of the Yeast genome. The occupancy score is given by a re-scaling function of the complexity value $f_c$ ($y$-axis) where the highest correlation value is 1 and the lowest 0. In the case of the Kaplan model, $f_c$ is the score calculated by the model~\cite{kaplan} itself which retrieves probability values between 0 and 1. Other cases not shown (e.g. entropy rate or Compress on RY or SW) yielded no significant results. Magenta and pink (bright colours) signify measures of algorithmic complexity; the information-theoretic based measures are in dark grey. This segment in chromosome 14 is the longest continuous segment for which in vivo and in vitro nucleosomal positioning values exist and is thus the reason this segment is recurrently used across several papers. When integrating more regions the errors accumulated due to large gaps of missing values produced results impossible to compare thereby forcing us to constrain the experiment to the this segment.}
  \label{topallplots}
\end{figure*}

In Fig.~\ref{topallplots} it was puzzling to find that the Kaplan model correlated less strongly than GC content alone for in vivo data, suggesting that the model assigns greater weight to $k$-mer information than to GC content for these extreme cases, given that we had known that the Kaplan model was mostly driven by GC content (Fig.~\ref{plots} middle). The box plot for the Kaplan model indicates that the model does not work as well for extreme sequences of high occupancy, with an average of $0.6$ where the maximum over the segments on which these nucleosome regions are contained reaches an average correlation of $\sim 0.85$ (in terms of occupancy), as shown in Fig.~\ref{plots} for in vitro data. This means that these high occupancy sequences are on the outer border of the standard deviation in terms of accuracy in the Kaplan model.

The best model is the one that best separates the highest from the lowest occupancy, and therefore is clearly Kaplan's model. Except for information-theoretic indices (Entropy and Compress), all algorithmic complexity indices were found to be informative of high and low occupancy. Moreover, all algorithmic complexity measures display a slight reduction in accuracy in vivo vs. in vitro, as is consistent with the limitations of current models such as Kaplan's. All but the Kaplan model are, however, training-free measures, in the sense that they do not contain any $k$-mer information related to high or low occupancy and thus are naive indices. Yet all algorithmic complexity measures were informative to different extents, with CTM and BDM performing best and LD performing worst, LD displaying inverted values for high and low occupancy as theoretically expected (because LD assigns low LD to high algorithmic complexity)~\cite{computability}. Also of note is the fact that CTM and BDM applied to the RY transformation were informative of high vs. low occupancy, thereby revealing a signal different from GC content that models such as Kaplan's partially capture in their encoded $k$-mer information. Interestingly, we found that GC content alone outperforms the Kaplan model for top highest and bottom lowest occupancies, both for in vitro and in vivo data, using 300K bps from three different Yeast chromosomal regions, even though the Kaplan model can outperform GC content by a small but significant fraction in some regions, such as chromosome 14. The three regions were taken from chromosome 3 (positions 100K to 200K), chromosome 8 (positions 100K to 200K) and chromosome 14 (147K to 247K).

Lossless compression was the worst behaved, showing how CTM and BDM outperform what is usually used as an estimator of algorithmic complexity~\cite{smalldata,emergence,bdm}. Unlike entropy alone, however, lossless compression does take into consideration sequence repetitions, averaging over all $k$-mers up to the compression algorithm sliding window length. The results thus indicate that averaging over all sequence motifs---both informative and not---deletes all advantages, thereby justifying specific knowledge-driven $k$-mer approaches introduced in models such as Segal's and Kaplan's.

\section{Conclusions}

Current gold standard prediction methods for nucleosome location highly correlate with GC content and require extensive (pre-)training to refine what GC content can achieve. Here we have gone beyond previous attempts to connect and apply measures of complexity to structural and functional properties of genomic DNA, specifically in the highly active and open challenge of nucleosome occupancy in molecular biology.

\begin{figure*}
  \centering
\includegraphics[width=16cm]{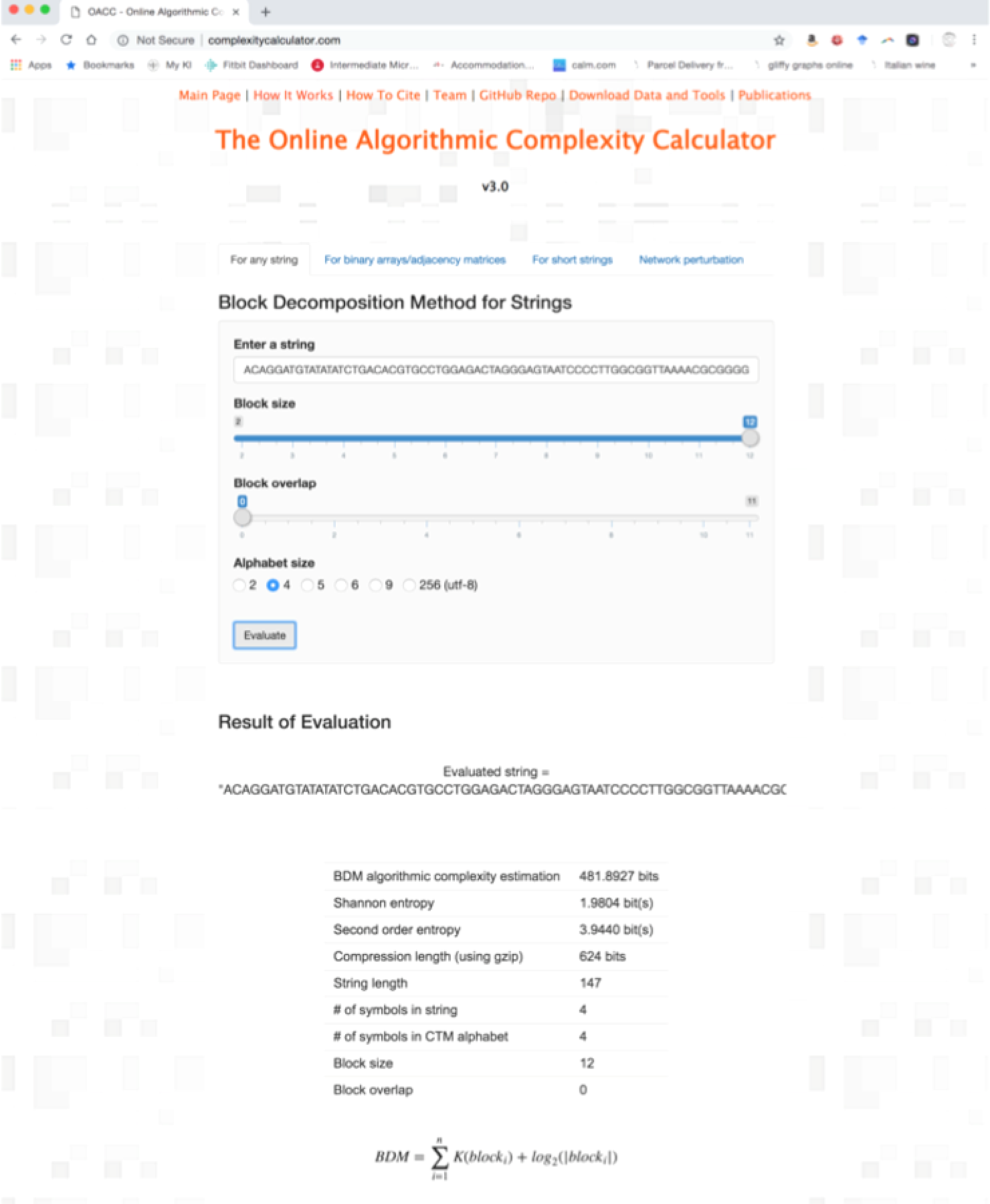}
\caption{Evaluation of nucleosome 601 DNA sequence in the Online Algorithmic Complexity Calculator freely available at \url{http://complexitycalculator.com/}. To reproduce scores between 0 and 1 as reported in all the results, suffices to rescale all values between 0 and 1.}
  \label{oacc}
\end{figure*}

While more investigation is needed these first experiments strongly suggest, and we report, that:

\begin{myenumerate2}
\item Algorithmic measures such as CTM and BDM of DNA sequences are informative of nucleosome occupancy/positioning. This is especially true for:
\begin{myenumerate2}
\item Short DNA sequences (Fig.~\ref{3a}), shorter than 5K bps on which, according to Kaplan et al. their model should not be used for.
\item Sequences that are experimentally shown to have very high (top 2\%) nucleosome occupancy (Fig.~\ref{topallplots}). For example, based on Fig. 4 CTM, CTM-SW, BDM and BDM-SW occupancy sequence complexity-based scores outperform the Kaplan model for predicting binding to sequences that are experimentally shown to be ones with very high-occupancy (top 2\%). In particular, we have a better separation between high and low occupancy for BDM vs. entropy and better correlation values for BDM vs. Kaplan: $\sim$ 0.82 (BDM) vs. $\sim$ 0.6 (Kaplan) for in vitro and $\sim$ 0.82 (BDM) vs. $\sim$ 0.43 (Kaplan) for in vivo. This represents a significant advancement in predicting high nucleosome occupancy using a training-free approach. These sequences have been reported to have particular biological significance (see~\cite{tillo}).
\item Sequences in which GC content may not be as informative. Because, unlike $k$-mer frequency-based scores, BDM does not trivially correlate with GC content as shown in Fig.~\ref{3a}. In contrast, we know that the correlation of the Kaplan model with GC content is very high (overall $\sim 0.90$ Pearson correlation based on Fig.~\ref{plots}).  
\end{myenumerate2}
\item Computational biologists can estimate CTM and BDM values for candidate nucleosomal DNA sequences of any length using an online complexity calculator, \url{http://complexitycalculator.com} and following these steps:
\begin{myenumerate2}
    \item[Step 0)] Chunk the DNA sequence into subsequences of desired sliding window length,
    \item[Step 1)] Introduce each DNA sequence in the calculator field (see Fig.~\ref{oacc} in the Sup. Mat.),
	\item[Step 2)] Retrieve the value for each query,
    \item[Step 3)] The ordered time series of CTM/BDM values is the score function. Lowest values are more likely to signal a nucleosome centre according to the results in this paper.
\end{myenumerate2}
Source code to perform these calculations without querying the website is also available online written in R and easily accessible through the acss package fully documented at:
\url{https://cran.r-project.org/web/packages/acss/acss.pdf}

\end{myenumerate2}

All the above points mean that CTM and BDM as training-free measures and GC content more independent (Fig.~\ref{3a}) may cover domains previously left uncovered by the Kaplan model, and that these new indices can complement current protocols making it possible to combine these measures with  the Kaplan method to produce even more accurate predictions.

%These training-free measures reveal that there is more structure to nucleosome occupancy than GC content, and potentially to $k$-mer structure as well (e.g. non-AT-based mers), based on the correlations found in RY transformations indicative of low versus high occupancy. 

%What the results imply is that even computable measures such as entropy do exploit $k$-mer information without explicit knowledge about them, and more sophisticated measures, such as lossless compression, BDM and LD can build upon that knowledge.

%The results raise the question of whether GC content is the cause or the effect of algorithmic information signals contained in the DNA sequence.

A direction for future research suggested by our work is the exploration of the use of these complexity indices to complement current machine learning approaches for reducing the feature space, by, e.g., determining which $k$-mers are more and less informative, and thereby ensuring better prediction results. Further investigation of the indices application to high nucleosome occupancy may also be required using some other gold standard organism such as C. elegans on which in vitro and in vivo nucleosomal data is comprehensive.

Another direction to explore is an extensive investigation of the possible use of genomic profiling for other types of structural and functional properties of DNA, with a view to contributing to, e.g., HiC techniques or protein encoding/promoter/enhancer region detection, and to furthering our understanding of the effect of extending the alphabet transformation of a sequence to epigenetics.

\newpage

%We should use \begin{materials}
%\end{materials}

\section*{Supplementary Material}

\section{Indices of Information and of Algorithmic Complexity}

Here we describe alternative measures to explore correlations from an information-theoretic and algorithmic (hence causal) complexity perspective.

\subsection{Shannon Entropy}

Central to information theory is the concept of Shannon's entropyy, which quantifies the average number of bits needed to store or communicate a message. Entropy determines that one cannot store (and therefore communicate) a symbol with $n$ different symbols in less than $\log(n)$ bits. In this sense, Entropy determines a lower limit below which no message can be further compressed, not even in principle. Another application (or interpretation) of Shannon's information theory is as a measure for quantifying the \emph{uncertainty} involved in predicting the value of a random variable. 

Shannon defined the Entropy $?$ of a discrete random variable $X$ with possible values ${x_1, \dots, x_n}$ and probability distribution $P(X)$ as:

$$H(X)=-\sum_{i=1}^n P(x_i) \log_2 P(x_i)$$

\noindent where if $P(x_i) = 0$ for some $i$, the value of the corresponding summand 0 $log_2(0)$ is taken to be 0.

\subsubsection{Entropy Rate}

The function $R$ gives what is variously denominated as rate or block Entropy and is Shannon Entropy over blocks or subsequences of $X$ of length $b$. That is, 

$$H_R(X)=\min_{b=1}^{b=|X|} H(X_b)$$

If the sequence is not statistically random, then $H_R(X)$ will reach a low value for some $b$, and if random, then it will be maximally entropic for all blocks $b$. $H_R(X)$ is computationally intractable as a function of sequence size, and typically upper bounds are realistically calculated for a fixed value of $b$ (e.g. a window length). Notice that, as discussed in the main text, having maximal Entropy does not by any means imply algorithmic randomness (c.f.~\ref{algo}).

\subsection{Lossless compression algorithms}

Two widely used lossless compression algorithms were employed. On the one hand, Bzip2 is a lossless compression method that uses several layers of compression techniques stacked one on top of the other, including Run-length encoding (RLE), Burrows?Wheeler transform (BWT), Move to Front (MTF) transform, and Huffman coding, among other sequential transformations. Bzip2 compresses more effectively than LZW, LZ77 and Deflate, but is considerably slower.

On the other hand, \textit{Compress} is a lossless compression algorithm based on the LZW compression algorithm. Lempel?Ziv?Welch (LZW) is a lossless data compression algorithm created by Abraham Lempel, Jacob Ziv, and Terry Welch, and is considered universal for an infinite sliding window (in practice the sliding window is bounded by memory or choice). It is considered \textit{universal} in the sense of Shannon Entropy, meaning that it approximates the Entropy rate of the source (an input in the form of a file/sequence). It is the algorithm of the widely used Unix file compression utility `Compress', and is currently in the international public domain.

\subsection{Measures of Algorithmic Complexity}
\label{algo}

A binary sequence $s$ is said to be random if its Kolmogorov complexity $C(s)$ is at least twice its length. It is a measure of the computational resources needed to specify the object. Formally, 

$$
C(s)=\min\{|p| : T(p)=s\}
$$

\noindent where $p$ is a program that outputs $s$ running on a universal Turing machine $T$. $C$ as a function taking $s$ to the length of the shortest computer program that produces $s$ is semi-computable and upper bound estimations are possible. The measure is today the accepted mathematical definition of randomness, among other reasons because it has been proven to be mathematically robust by virtue of the fact that several independent definitions converge to it.

The invariance theorem guarantees that complexity values will only diverge by a constant  (e.g. the length of a compiler, a translation program between $T_1$ and $T_2$) and will converge at the limit. Formally,

$$
|C(s)_{T_1}-C(s)_{T_2}|<c
$$

\subsubsection{Lossless Compression as Approximation to $C$}

Lossless compression is traditionally the method of choice when a measure of algorithmic content related to Kolmogorov-Chaitin complexity $C$ is needed. The Kolmogorov-Chaitin complexity of a sequence $s$ is defined as the length of the shortest computer program $p$ that outputs $s$ running on a reference universal Turing machine $T$. While lossless compression is equivalent to algorithmic complexity, actual implementations of lossless compression (e.g. Compress) are heavily based upon Entropy rate estimations~\cite{emergence,bdm} that mostly deal with statistical repetitions or $k$-mers of up to a window length size $L$, such that $k\leq L$.

\subsubsection{Algorithmic Probability as Approximation to $C$}

Another approach consists in making estimations by way of a related measure, \textit{Algorithmic Probability}~\cite{delahayezenil,zenild5}. The Algorithmic Probability of a sequence $s$ is the probability that $s$ is produced by a random computer program $p$ when running on a reference Turing machine $T$. Both algorithmic complexity and Algorithmic Probability rely on $T$, but invariance theorems for both guarantee that the choice of $T$ is asymptotically negligible. 

One way to minimise the impact of the choice of $T$ is to average across a large set of different Turing machines all of the same size. The chief advantage of algorithmic indices is that causal signals in a sequence may escape entropic measures if they do not produce statistical regularities. And it has been the case that increasing the length of $k$ in $k$-nucleotide models of structural properties of DNA have not returned more than a marginal advantage. 

The Algorithmic Probability~\cite{solomonoff} (also known as Levin's semi-measure~\cite{levin}) of a sequence $s$ is a measure that describes the expected probability of a random program $p$ running on a universal prefix-free Turing machine $T$ producing $s$. Formally,

$$
m(s)=\sum_{p:T(p)=s} 1/{2^{|p|}}
$$

The Coding theorem beautifully connects $C(s)$ and $m(s)$:

$$
C(s) \sim -\log m(s)
$$

\subsubsection{Bennett's Logical Depth}

Another measure of great interest is \textit{logical depth}~\cite{bennett}. The logical depth (LD) of a sequence $s$ is the shortest time logged by the shortest programs $p_i$ that produce $s$ when running on a universal reference Turing machine. In other words, just as algorithmic complexity is associated with lossless compression, LD can be associated with the shortest time that a Turing machine takes to decompress the sequence $s$ from its shortest computer description. A multiplicative invariance theorem for LD has also been proven~\cite{bennett}. Estimations of Algorithmic Probability and logical depth of DNA sequences were performed as determined in~\cite{delahayezenil,zenild5}. 

Unlike algorithmic (Kolmogorov-Chaitin) complexity $C$, logical depth is a measure related to `structure' rather than randomness. LD can be identified with biological complexity~\cite{bennett2,collier} and is therefore of great interest when comparing different genomic regions.

\subsection{Measures Based on Algorithmic Probability and on Logical Depth}

The \textit{Coding theorem method} (or simply CTM) is a method~\cite{delahayezenil,zenild5} rooted in the relation between $C(s)$ and $m(s)$ specified by Algorithmic Probability, that is, between frequency of production of a sequence from a random program and its Kolmogorov complexity as described by Algorithmic Probability. Essentially, it uses the fact that the more frequent a sequence the lower its Kolmogorov complexity, and sequences of lower frequency have higher Kolmogorov complexity. Unlike algorithms for lossless compression, the Algorithmic Probability approach not only produces estimations of $C$ for sequences with statistical regularities, but it is deeply rooted in a computational model of Algorithmic Probability, and therefore, unlike lossless compression, has the potential to identify regularities that are not statistical (e.g. a sequence such as 1234...), that is, sequences with high Entropy or no statistical regularities but low algorithmic complexity~\cite{emergence,smalldata}. 

Let $(n,m)$ be the space of all $n$-state $m$-symbol Turing machines, $n,m > 1$ and $s$ a sequence, then:

$$D(n,m)(s) = \frac{|\{T \in (n,m): T \textit{ produces $s$}\}|}{|\{T \in (n,m)\}|}$$

\noindent where $T$ is a standard Turing machine as defined in the Busy Beaver problem by Rad\'o~\cite{rado} with 4 symbols (in preparation for the calculation of the DNA alphabet size).

Then using the relation established by the Coding theorem, we have:

$$CTM(s) = -\log_2(D(n,m)(s))$$

That is, the more frequently a sequence is produced the lower its Kolmogorov complexity, and vice versa. CTM is an upper bound estimation of Kologorov-Chaitin complexity.

From CTM, a measure of Logical Depth can also be estimated--as the computing time that the shortest Turing machine (i.e. the first in the quasi-lexicographic order) takes to produce its output $s$ upon halting. CTM thus produces both an empirical distribution of sequences up to a certain size, and an LD estimation based on the same computational model. 

Because CTM is computationally very expensive (equivalent to the Busy Beaver problem~\cite{rado}), only short sequences (currently only up to length $k=12$) have associated estimations of their algorithmic complexity. To approximate the complexity of genomic DNA sequences up to length $k=12$, we calculated $D(5,4)(s)$, from which $CTM(s)$ was approximated. 

To calculate the Algorithmic Probability of a DNA sequence (e.g. the sliding window of length 147 nt) we produced an empirical Algorithmic Probability distribution from $(5,4)$ to compare with by running a sample of 325\,433\,427\,739 Turing machines with up to 5 states and 4 symbols (the number of nucleotides in a DNA sequence) with empty input (as required by Algorithmic Probability). The resulting distribution came from 325\,378\,582\,327 non-unique sequences (after removal of those sequences only produced by 5 or fewer machines/programs). 

\subsection{Relation of BDM to Shannon Entropy and GC Content}

The Block Decomposition Method (BDM) is a divide-and-conquer method that can be applied to longer sequences on which local approximations of $C(s)$ using CTM can be averaged, thereby extending the range of application of CTM. Formally,

$$BDM(s,k) = \displaystyle \sum_{s_k}
\text{log}(n) + CTM(r)$$

\noindent where the set of subsequences $s_k$ is composed of the pairs $(r,n)$, where $r$ is an element of the decomposition of sequence $s$ of size $k$, and $n$ the multiplicity of each subsequence of length $k$. $BDM(s)$ is a computable approximation from below to the algorithmic information complexity of $s$, $C(s)$. BDM approximations to $C$ improve with smaller departures (i.e. longer $k$-mers) from the Coding Theorem method. When $k$ decreases in size, however, we have shown~\cite{bdm} that BDM approximates the Shannon Entropy of $s$ for the chosen $k$-mer distribution. In this sense, BDM is a hybrid complexity measure that in the `worst case' behaves like Shannon Entropy and in the best approximates $C$. We have also shown that BDM is robust when instead of partitioning a sequence, overlapping subsequences are used, but this latter method tends to over-fit the value of the resultant complexity of the original sequence that was broken into $k$-mers.

%%%%%%%%%%%%%%%%%% TABLES SECTION

\begin{table}
\begin{center}
\caption{Spearman correlations between complexity indices with in vivo and in vitro experimental nucleosome occupancy data from position 187\,001 bp to 207\,000 bp on the 14th Yeast chromosome}\label{infomeasures}
\begin{tabular}{c|c|c}
& \textbf{in vitro} & \textbf{in vivo}\\
\hline
\hline
\textbf{in vitro} & 1 & 0.5\\
\hline
\textbf{in vivo} & 0.5 & 1\\
\hline
\textbf{GC content} & 0.684 & 0.26\\
\hline
\textbf{LD} & -0.29 & -0.23\\
\hline
\textbf{Entropy} & 0.588 & 0.291 \\ 
\hline
\textbf{BDM} & 0.483 & 0.322\\
\hline
\textbf{Compress} & 0.215 & 0.178
\end{tabular}
\end{center}
\end{table}

\begin{table}
\begin{center}
{\footnotesize
\caption{Spearman correlation values of complexity score functions vs. the Wedge dinucleotide model prediction of DNA curvature on 20 synthetically generated DNA sequences depicted in Table~\ref{sequences}}\label{heatmap}
{\tiny
\begin{tabular}{c|c|c|c|c|c|c|c}
& \textbf{GC} & \textbf{Entropy} & \textbf{Entropy} & \textbf{Compress} & \textbf{BZip2} & \textbf{BDM} & \textbf{LD}\\
& \textbf{content} &  &  \textbf{rate} (4) &  &  &  &  \\
\hline
\hline
\textit{rho} & -0.45 & -0.44 & -0.57 & -0.58 & -0.45 & -0.57 & 0.65 \\
\hline
\textit{p} & 0.047 & 0.051  & 0.0094 & 0.0079 & 0.048 & 0.0083  & 0.0019  \\
 \end{tabular}
 }
 }
\end{center}
\end{table}

\begin{table}
\begin{center}
{\footnotesize
\caption{Distance in number of nucleotides to local minimum in all cases except LD (for which local maximum was taken) and for GC content for which min/max within a window of 73 nts around the nucleosomal centre was calculated. In all cases, pseudo-randomly generated sequences with the same GC content as the mean of the GC content of the next nucleosomal neighbour was inserted because the purpose of the experiment is for a GC content test for nucleosome location to fail. However, even in cases when GC content is not informative, BDM is able to locate the nucleosome centre in a high number of cases and within an accuracy of less than 16 nts on average. Entropy is off by around 20 nts, lossless compression  by more than 26 nts, GC content  by more than 30 nts and LD was actually correlated with areas outside the centre, i.e. marking the opposite position to the centres and thus negatively correlated with centre location.}
\label{nucleosomedistances}
{\tiny
 \begin{tabular}{c|c|c|c|c|c|c|c}
  \multicolumn{8}{c}{\textbf{BDM}}\\
\textbf{601} & \textbf{603} & \textbf{605} & \textbf{5Sr DNA} & \textbf{pGub} & \textbf{chicken} $\beta -$ &\\
&&&&& \textbf{globulin} & &\\
\hline
5 & 19 & 13 & 59 & 25 & 6 & & \\
\hline
\hline
\textbf{msat} & \textbf{CAG} & \textbf{TATA} & \textbf{CA} & \textbf{NoSecs} & \textbf{TGGA} & \textbf{TGA} & \textbf{BadSecs}\\
\hline
42 & 2 & 1 & 1 & 29 & 1 & 1 & 7\\
\end{tabular}
}
}
\end{center}
\end{table}

\begin{table}
\caption{\label{sequences}The 20 short DNA sequences artificially generated covering a wide range of patterns and regularities used to find informative measures of DNA curvature.}
\begin{center}
\begin{tabular}{c|c|c}
{\footnotesize AAAAAAAAAAAA}& {\footnotesize ATATATATATAT}& {\footnotesize AAAAAATTTTTT}\\
\hline
{\footnotesize AAAAAAAAATAA}&{\footnotesize AAAAAAAACAAT}&{\footnotesize AAGATCTACACT}\\
\hline
{\footnotesize ATAGAACGCTCC}&{\footnotesize ACCTATGAAAGC}&{\footnotesize TAGGCGGCGGGC}\\
\hline
{\footnotesize TCGTTCGCGAAT}&{\footnotesize TGCACGTGTGGA}&{\footnotesize CTAAACACAATA}\\
\hline
{\footnotesize CTCTCAGGTCGT}&{\footnotesize CTCGTGGATATC}&{\footnotesize CCACGATCCCGT}\\
\hline
{\footnotesize GGCGGGGGGTGG}&{\footnotesize GGGGGGGCGGGC}&{\footnotesize GGGGGGCCCCCC}\\
\hline
{\footnotesize GCGCGCGCGCGC}&{\footnotesize GGGGGGGGGGGG}&
\end{tabular}
\end{center}
\end{table}

\begin{table}
\caption{\label{nucleosome}14 Experimental nucleosome sequences~\cite{heijden}. Only the first 6 have known dyads}
\begin{center}
{\footnotesize
%\vspace{-3cm}
\begin{tabular}{c|c|c}
\textbf{name} & \textbf{dyad} & \textbf{sequence}\\
& \textbf{position} &\\
\hline
601 & 74 & {\tiny ACAGGATGTATATATCTGACACGTGCCTGGAGACTAGGGAGTA}\\

&&{\tiny ATCCCCTTGGCGGTTAAAACGCGGGGGACAGCGCGTACGTGCG}\\

&&{\tiny TTTAAGCGGTGCTAGAGCTGTCTACGACCAATTGAGCGGCCTCG}\\

&&{\tiny GCACCGGGATTCTCCAG}\\
 \hline
603 & 154 & {\tiny CGAGACATACACGAATATGGCGTTTTCCTAGTACAAATCACCCCA}\\

&&{\tiny GCGTGACGCGTAAAATAATCGACACTCTCGGGTGCCCAGTTCGC}\\

&&{\tiny GCGCCCACCTACCGTGTGAAGTCGTCACTCGGGCTTCTAAGTACG}\\

&&{\tiny CTTAGGCCACGGTAGAGGGCAATCCAAGGCTAACCACCGTGCAT}\\

&&{\tiny CGATGTTGAAAGAGGCCCTCCGTCCTTATTACTTCAAGTCCCTGG}\\

&&{\tiny GGTACCGTTTC}\\
  \hline
605 &132&{\tiny TACTGGTTGGTGTGACAGATGCTCTAGATGGCGATACTGACAGG}\\

&&{\tiny TCAAGGTTCGGACGACGCGGGATATGGGGTGCCTATCGCACATT}\\

&&{\tiny GAGTGCGAGACCGGTCTAGATACGCTTAAACGACGTTACAACCC}\\

&&{\tiny TAGCCCCGTCGTTTTAGCCGCCCAAGGGTATTCAAGCTCGACGCT}\\

&&{\tiny AATCACCTATTGAGCCGGTATCCACCGTCACGACCATATTAATAG}\\

&&{\tiny GACACGCCG}\\
  \hline
5Sr DNA & 74, 92 & {\tiny AACGAATAACTTCCAGGGATTTATAAGCCGATGACGTCATAACAT}\\

&&{\tiny CCCTGACCCTTTAAATAGCTTAACTTTCATCAAGCAAGAGCCTAC}\\

&&{\tiny GACCATACCATGCTGAATATACCGGTTCTCGTCCGATCACCGAAG}\\

&&{\tiny TCAAGCAGCATAGGGCTCGGTTAGTACTTGGATGGGAGACCGCC}\\

&&{\tiny TGGGAATACCG}\\
  \hline
pGub & 84, 104& {\tiny GATCCTCTAGACGGAGGACAGTCCTCCGGTTACCTTCGAACCACGT}\\

&&{\tiny GGCCGTCTAGATGCTGACTCATTGTCGACACGCGTAGATCTGCTAG}\\

&&{\tiny CATCGATCCATGGACTAGTCTCGAGTTTAAAGATATCCAGCTGCCC}\\

&&{\tiny GGGAGGCCTTCGCGAAATATTGGTACCCCATGGAATCGAGGGATC}\\
  \hline
chicken $\beta$-&125& {\tiny CTGGTGTGCTGGGAGGAAGGACCCAACAGACCCAAGCTGTGGTC}\\

globulin&&{\tiny TCCTGCCTCACAGCAATGCAGAGTGCTGTGGTTTGGAATGTGTGA}\\

&&{\tiny GGGGCACCCAGCCTGGCGCGCGCTGTGCTCACAGCACTGGGGTG}\\

&&{\tiny AGCACAGGGTGCCATGCCCACACCGTGCATGGGGATGTATGGCGC}\\

&&{\tiny ACTCCGGTATAGAGCTGCAGAGCTGGGAATCGGGGGG}\\
  \hline
mouse minor &&{\tiny ATTTGTAGAACAGTGTATATCAATGAGCTACAATGAAAATCATGGA}\\
satellite&&{\tiny AAATGATAAAAACCACACTGTAGAACATATTAGATGAGTGAGTTA}\\

&&{\tiny CACTGAAAAACACATCCGTTGGAAACCGGCAT}\\
  \hline
CAG&& {\tiny AGCAGCAGCAGCAACAGTAGTAGAAGCAGCAGCACTAACGACAG}\\

&&{\tiny CACAGCAGTAGCAGTAATAGAAGCAGCAGCAGCAGCAGTAGCAG}\\

&&{\tiny TAGCAGCAGCAGCAGCAGCAATTTCAACAACAGCAGCAGCAGCT}\\
  \hline
TATA&& {\tiny AGGTCTATAAGCGTCTATAAGCGTCTATGAACGTCTATAAACGTCT}\\

&&{\tiny ATAAACGCCTATAAACGCCTATAAACGCCTATACAAGCCTATAAAC}\\

&&{\tiny GCCTATACACGTCTATGCACGACTATACACGTCT}\\
  \hline
CA&& {\tiny GAGAGTAACACAGGCACAGGTGTGGAGAGTAACACAGGCACAG}\\

&&{\tiny GTGTGGGAGAGTGACACACAGGCACAGGTGAGGAGAGTACACA}\\

&&{\tiny CAGGCACAGGTGTGGAGAGCACACACAGGTGCGGAGAG}\\
  \hline
NoSecs&& {\tiny GGGCTGTAGAATCTGATGGAGGTGTAGGATGGATGGACAGTATGA}\\

&&{\tiny CAAAAGGGTACTAGCCTGGGACAGCAGGATTGGTGGAAAGGTTA}\\

&&{\tiny CAGGCAGGCCCAGCAGGCTCGGACGCTGTATAGAG}\\
  \hline
TGGA&& {\tiny AGATGGATGGATGATGGATGGATGATGGATAGATGGATGATGGAT}\\

&&{\tiny GGATGGATGATGATGGATGAATAGATGGATGGATGGATGATGGAT}\\

&&{\tiny GGATGGACGATGGATGGATAGATGGATGGATGG}\\
  \hline
TGA && {\tiny ATAGATGGATGAGTGGATGGATGGGTGGATGGATAGATGGGTGG}\\

&&{\tiny ATGGGTGGATGGGTGGATGGATGATGGATGGATGAGTGGATGGA}\\

&&{\tiny TGGATGGATGGGTGGATGGGTGGACGG}\\
\hline
BadSecs && {\tiny TCTAGAGTGTACAACTATCTACCCTGTAGGCATCAAGTCTATTTCGG}\\

&&{\tiny TAATCACTGCAGTTGCATCATTTCGATACGTTGCTCTTGCTTCGCTAG}\\

&&{\tiny CAACGGACGATCGTACAAGCAC}
 \end{tabular}
 }
\end{center}
%\vspace{-.7cm}
\end{table}

\begin{thebibliography}{100}

\bibitem{tanmoy} Tanmoy T. (2010). Characterization of the RNA content of chromatin. \emph{Genome Res.} 20 (7): 899?907. 

\bibitem{reece} Reece J, Campbell N (2006). \emph{Biology}. San Francisco: Benjamin Cummings.

\bibitem{Yeast} Tillo D, Hughes TR (2009) {G}+{C} content dominates intrinsic nucleosome occupancy.
\newblock {\em BMC bioinformatics} 10(1):1.

\bibitem{tillo} Tillo D, Kaplan N, Moore IK, Fondufe-Mittendorf Y, Gossett AJ, Field Y, et al. (2010) High Nucleosome Occupancy Is Encoded at Human Regulatory Sequences. PLoS ONE 5(2): e9129.

\bibitem{struhl} Struhl K, Segal E (2013) Determinants of nucleosome positioning.
\newblock {\em Nat Struct Mol Biol.} Mar;20(3):267-73.

%1
\bibitem{changu}
Gu C et~al. (2015) {D}{N}{A} structural correlation in short and long ranges.
\newblock {\em The Journal of Physical Chemistry B} 119(44):13980--13990.

%2
\bibitem{schep}
Schep AN et~al. (2015) Structured nucleosome fingerprints enable
  high-resolution mapping of chromatin architecture within regulatory regions.
\newblock {\em Genome research} 25(11):1757--1770.

%3
\bibitem{kaplan} Kaplan N et~al. (2009) The {D}{N}{A}-encoded nucleosome organisation of a eukaryotic genome.
\newblock {\em Nature} 458(7236):362--366.

%4
\bibitem{gracey}
Gracey LE et~al. (2010) An in vitro-identified high-affinity
  nucleosome-positioning signal is capable of transiently positioning a
  nucleosome in vivo.
\newblock {\em Epigenetics \& chromatin} 3(1):1.

%5
\bibitem{rivals}
Rivals E, Delahaye JP, Dauchet M, Delgrange O (1996) Compression and genetic sequence analysis.
\newblock {\em Biochimie} 78:315--322.

%6
\bibitem{vitanyi2}
Cilibrasi R, Vit{\'a}nyi PM (2005) Clustering by compression.
\newblock {\em IEEE Transactions on Information theory} 51(4):1523--1545.

%7
\bibitem{pratas} Pratas D, Pinho, AJ (2017)
On the Approximation of the Kolmogorov Complexity for DNA. In: Alexandre L., Salvador Sánchez J., Rodrigues J. (eds) Pattern Recognition and Image Analysis. IbPRIA 2017. Lecture Notes in Computer Science, vol 10255. Springer.

\bibitem{vitanyi}
Li M, Chen X, Li X, Ma B, Vit{\'a}nyi PM (2004) The similarity metric.
\newblock {\em IEEE transactions on Information Theory} 50(12):3250--3264.



\bibitem{utro} Utro F, Di Benedetto V, Corona DVF and Giancarlo R (2015) The intrinsic combinatorial organization and information theoretic content of a sequence are correlated to the DNA encoded nucleosome organization of eukaryotic genomes, \newblock {\em Bioinformatics}, 1--8.

%7.5
\bibitem{emergence}
Zenil H, Badillo L,  Hern\'andez-Orozco, Hern\'andez-Quiroz F (2018) Coding-theorem Like Behaviour and Emergence of the Universal Distribution from Resource-bounded Algorithmic Probability,  
\newblock {\em International Journal of Parallel Emergent and Distributed Systems} (accepted)

%8
\bibitem{pozzoli}
Pozzoli U et~al. (2008) Both selective and neutral processes drive {G}{C}
  content evolution in the human genome.
\newblock {\em BMC evolutionary biology} 8(1):1.

\bibitem{galtier}
Galtier N, Piganeau G, Mouchiroud D, Duret L (2001) {G}{C}-content evolution in
  mammalian genomes: the biased gene conversion hypothesis.
\newblock {\em Genetics} 159(2):907--911.

%11
\bibitem{minary}
Minary P, Levitt M (2014) Training-free atomistic prediction of nucleosome
  occupancy.
\newblock {\em Proceedings of the National Academy of Sciences}
  111(17):6293--6298.

\newblock {\em Lecture Notes in Computer Science book series (LNCS)}, vol. 10255:pp 259--266.

%12
\bibitem{cui}
Cui F, Zhurkin VB (2010) Structure-based analysis of {D}{N}{A} sequence
  patterns guiding nucleosome positioning in vitro.
\newblock {\em Journal of Biomolecular Structure and Dynamics} 27(6):821--841.

%13
\bibitem{trifonov} Trifonov, E. N., and Sussman, J. L. (1980) The pitch of chromatin DNA is reflected in its nucleotide sequence. \newblock {\em Proc. Natl. Acad. Sci. USA} 77, pp. 3816--3820.

%14
\bibitem{segal}
Segal E et~al. (2006) A genomic code for nucleosome positioning.
\newblock {\em Nature} 442(7104):772--778.

%15
\bibitem{kelley}
Kelley DR, Snoek J, Rinn JL (2016) Basset: Learning the regulatory code of the
  accessible genome with deep convolutional neural networks.
\newblock {\em Genome research}, 26(7):990--9.

%16
\bibitem{lee}
Lee W et~al. (2007) A high-resolution atlas of nucleosome occupancy in yeast.
\newblock {\em Nature genetics} 39(10):1235--1244.

%17
\bibitem{kanhere}
Kanhere A, Bansal M (2003) An assessment of three dinucleotide parameters to
  predict {D}{N}{A} curvature by quantitative comparison with experimental
  data.
\newblock {\em Nucleic acids research} 31(10):2647--2658.

%18
\bibitem{ulanovsky}
Ulanovsky LE, Trifonov EN (1986) Estimation of wedge components in curved
  {D}{N}{A}.
\newblock {\em Nature} 326(6114):720--722.

%19
\bibitem{burkhoff}
Burkhoff AM, Tullius TD (1988) Structural details of an adenine tract that does
  not cause {D}{N}{A} to bend.
\newblock {\em Nature} 331:455--457.

%20
\bibitem{donald}
Crothers DM, Haran TE, Nadeau JG (1990) Intrinsically bent {D}{N}{A}.
\newblock {\em J. Biol. Chem} 265(13):7093--7096.

%21
\bibitem{sinden}
Sinden RR (2012) {\em {D}{N}{A} structure and function}.
\newblock (Elsevier).

%22
\bibitem{heijden}
van~der Heijden T, van Vugt JJ, Logie C, van Noort J (2012) Sequence-based
  prediction of single nucleosome positioning and genome-wide nucleosome
  occupancy.
\newblock {\em Proceedings of the National Academy of Sciences}
  109(38):E2514--E2522.

%23
\bibitem{bennett}
Bennett CH (1995) Logical depth and physical complexity.
\newblock {\em The Universal {T}uring Machine, A Half-Century Survey} pp.
  207--235.

%24
\bibitem{bennett2}
Bennett CH (1993) Dissipation, information, computational complexity and the
  definition of organisation in {\em Santa Fe Institute Studies in the Sciences
  of Complexity -Proceedings Volume-}.
\newblock (Addison-Wesley Publishing Company), Vol.{}~1, pp. 215--215.

%25
\bibitem{alife}
Hern{\'a}ndez-Orozco S, Hern{\'a}ndez-Quiroz F, Zenil H (2016) The limits of decidable states on open-ended evolution and emergence in {\em ALIFE Conference}.
\newblock (MIT Press).

%26
%\bibitem{scirep}
%Adams AM, Zenil H, Davies PC, Walker SI (2016) Formal definitions of unbounded evolution and innovation reveal universal mechanisms for open-ended evolution in dynamical systems.
%\newblock {\em Scientific Reports (accepted)}.

%27
\bibitem{bdm}
Zenil H, Soler-Toscano F, Kiani NA, Hern{\'a}ndez-Orozco S, Rueda-Toicen A
  (2016) A decomposition method for global evaluation of {S}hannon Entropy and
  local estimations of algorithmic complexity.
\newblock {\em arXiv preprint arXiv:1609.00110}.

%28
\bibitem{solomonoff}
Solomonoff RJ (1964) A formal theory of inductive inference. parts i and ii.
\newblock {\em Information and control} 7(1):1--22 and 224--254.

%29
\bibitem{levin}
Levin LA (1974) Laws of information conservation (nongrowth) and aspects of the
  foundation of probability theory.
\newblock {\em Problemy Peredachi Informatsii} 10(3):30--35.

%30
\bibitem{kolmogorov}
Kolmogorov AN (1968) Three approaches to the quantitative definition of
  information.
\newblock {\em International Journal of Computer Mathematics} 2(1-4):157--168.

%31
\bibitem{chaitin}
Chaitin GJ (1969) On the length of programs for computing finite binary
  sequences: statistical considerations.
\newblock {\em Journal of the ACM (JACM)} 16(1):145--159.

%32
\bibitem{delahayezenil}
Delahaye JP, Zenil H (2012) Numerical evaluation of algorithmic complexity for
  short strings: A glance into the innermost structure of randomness.
\newblock {\em Applied Mathematics and Computation} 219(1):63--77.

%33
\bibitem{zenild5}
Soler-Toscano F, Zenil H, Delahaye JP, Gauvrit N (2014) Calculating kolmogorov
  complexity from the output frequency distributions of small {T}uring
  machines.
\newblock {\em PloS one} 9(5):e96223.

%33.5

\bibitem{computability}
Soler-Toscano F, Zenil H, Delahaye JP, Gauvrit N (2013) Correspondence and Independence of Numerical Evaluations of Algorithmic Information Measures
\newblock{\em Computability}, vol. 2, no. 2, pp 125--140.

%34
\bibitem{klug}
Klug A, Rhodes D, Smith J, Finch J, Thomas J (1980) A low resolution structure
  for the histone core of the nucleosome.
\newblock {\em Nature} 287(5782):509--516.

%35
\bibitem{gaykalova}
Gaykalova DA et~al. (2011) A polar barrier to transcription can be circumvented
  by remodeler-induced nucleosome translocation.
\newblock {\em Nucleic acids research} p. gkq1273.

%36
\bibitem{smalldata}
Zenil H (2013) Algorithmic data analytics, small data matters and correlation versus causation in {\em Computability of the World? Philosophy and Science
  in the Age of Big Data}, eds.{} Pietsch W, Wernecke J, Ott M.
\newblock (Springer Verlag). In press.

%37
\bibitem{collier}
Collier JD (1998) Information increase in biological systems: how does
  adaptation fit? in {\em Evolutionary systems}.
\newblock (Springer), pp. 129--139.

%38
\bibitem{rado}
Rado T (1962) On non-computable functions.
\newblock {\em Bell System Technical Journal} 41(3):877--884.

%\bibitem{milosavljevic}
%Milosavljevi{\'c} A (1995) Discovering dependencies via algorithmic mutual
%  information: A case study in {D}{N}{A} sequence comparisons.
%\newblock {\em Machine Learning} 21(1-2):35--50.


%\bibitem{trifonov}
%Trifonov EN (2010) Nucleosome positioning by sequence, state of the art and
%  apparent finale.
%\newblock {\em Journal of Biomolecular Structure and Dynamics} 27(6):741--746.


\end{thebibliography}
\end{document}